\documentclass[pre,twocolumn,aps,superscriptaddress,floats,showpacs]{revtex4}
\usepackage{amssymb}
\usepackage{graphicx}
\usepackage{amsmath}

\begin{document}

\title{On the interaction of the electromagnetic radiation with the breaking plasma waves}
\author{A.~V.~Panchenko}
\affiliation{Moscow Institute of Physics and Technology, Institutskii pereulok 9,
Dolgoprudnyi, Moscow region, 141700 Russia}
\author{T.~Zh.~Esirkepov}
\affiliation{Kansai Photon Science Institute, Japan Atomic Energy Agency, 8-1 Umemidai,
Kizugawa, Kyoto, 619-0215 Japan}
\author{A.~S.~Pirozhkov}
\affiliation{Kansai Photon Science Institute, Japan Atomic Energy Agency, 8-1 Umemidai,
Kizugawa, Kyoto, 619-0215 Japan}
\author{M. Kando}
\affiliation{Kansai Photon Science Institute, Japan Atomic Energy Agency, 8-1 Umemidai,
Kizugawa, Kyoto, 619-0215 Japan}
\author{F.~F.~Kamenets}
\affiliation{Moscow Institute of Physics and Technology, Institutskii pereulok 9,
Dolgoprudnyi, Moscow region, 141700 Russia}
\author{S.~V.~Bulanov}
\affiliation{Kansai Photon Science Institute, Japan Atomic Energy Agency, 8-1 Umemidai,
Kizugawa, Kyoto, 619-0215 Japan}

\begin{abstract}
An electromagnetic wave (EMW) interacting with the moving singularity of the
charged particle flux undergoes the reflection and absorption as well as
frequency change due to Doppler effect and nonlinearity. The singularity
corresponding to a caustic in plasma flow with inhomogeneous velocity can
arise during the breaking of the finite amplitude Langmuir waves due to
nonlinear effects. A systematic analysis of the wave-breaking regimes and
caustics formation is presented and the EMW reflection coefficients are
calculated.
\end{abstract}

\maketitle

\section{Introduction}

A strong interest has persisted in studying the electromagnetic wave (EMW)
interaction with the relativistic electron structures represented by
electron beams \cite{rel-beam}, Langmuir waves \cite{Photon
acceleration,KrSoob,Light intensification,Kando-2007,Pirozhkov-2007}
ionization fronts \cite{Ioniz}, relativistic solitons and vortices \cite%
{RelSol} in an underdense plasma, and by the oscillating electron layers at
a solid target surface \cite{OscMir,Cher,Naumova}.

The laser produced irradiance approaches 10$^{22}$ W/cm$^{2}$ \cite{1022},
which leads to ultrarelativistic plasma dynamics. By further increasing the
irradiance we shall see novel physical regimes such as the radiation
friction force dominated EMW-matter interaction \cite{RadF}. At irradiances
around \ 10$^{28}$ W/cm$^{2}$ \cite{H-E}, the focused light becomes so
strong that the nonlinear effects come into play, as predicted by quantum
electrodynamics, including the vacuum polarization and the electron-positron
pair creation from vacuum (see review articles \cite{MTB} and the
liteterature quoted therein). The studies of the EMW interaction with the
relativistic structures pave a way towards achieving such the irradiance.

Here we pay the main attention to the ``flying mirror'' concept \cite{Light
intensification} (see its further development and discussion in Refs. \cite%
{BEKT,MTB,RCNP}). The electromagnetic radiation source suggested in this
concept has unique advantages. It is robust since it is based on the
fundamental process of wave breaking. It enables the control of the output
pulse parameters such as frequency, duration, focusing, spectrum, etc.,
required by a wide range of applications. It allows so high up-shifting of
the laser radiation frequency, followed by focusing to a spot whith size
determined by the shortened wavelength, that the quantum electrodynamics
critical field (Schwinger limit) can be achieved with present-day laser
systems \cite{Light intensification}. 

\begin{figure}
\includegraphics[scale=1]{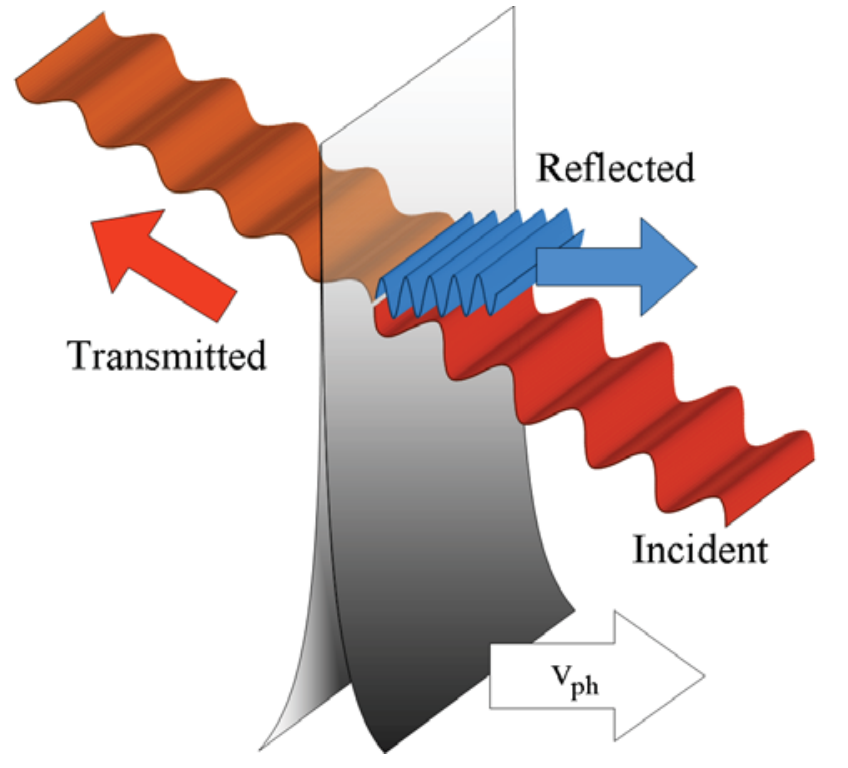}
\caption{(color online)
The ``flying mirror'' concept.
The pulse incident on the moving density
cusp is partially reflected.
The reflected pulse frequency is upshifted.}
\label{fig:FM}
\end{figure}

In the ``flying mirror'' concept, extremely high density thin electron
shells in the strongly nonlinear plasma wave, generated in the wake of an
ultra-short laser pulse, act as semi-transparent mirrors flying with a
velocity close to the speed of light, Fig. \ref{fig:FM}.
The mirrors reflect a
counter-propagating pulse, whose frequency is upshifted, duration is
shortened and power is increased. The obtained pulse power is proportional
to the Lorentz factor $\gamma _{ph}=(1-\beta _{ph}^{2})^{-1/2}$
corresponding to the velocity of the mirrors $v_{ph}=\beta _{ph}c$ while the
frequency and the pulse compression are proportional to the square of the
Lorentz factor.

A nonlinear wake wave with substantially large amplitude, at which the
Lorentz factor of electrons, $\gamma_{e}$, is greater than that of the wake
wave, $\gamma_{e}>\gamma _{ph}$, breaks and forms cusps in the electron
density distribution \cite{Akhiezer-Polovin}. It is the steepness of the
cusps that affords the efficient reflection of a portion of the
counter-propagating laser pulse. If the wake wave is far below the
wave-breaking threshold, the reflection is exponentially small. The
frequency of the reflected light is upshifted due to the double Doppler
effect, as predicted by Einstein \cite{Einstein}. In his paper, the problem
of the reflection of light from a mirror moving with constant velocity, $%
v_{M}=c\beta _{M}$, close to the speed of light, is solved as an example of
the use of relativistic transformations. The incident and the reflected wave
frequencies, $\omega _{0}$ and $\omega _{r}$, are related to each other as 
\begin{equation}
\omega _{r}=\omega _{0}\frac{1+\beta _{M}^{2}+2\beta _{M}\cos \theta _{0}}{%
1-\beta _{M}^{2}},  \label{eq.1-omr}
\end{equation}%
where $\theta _{0}$ is the wave incidence angle. The reflection angle, $%
\theta _{r}$, is defined by the expression 
\begin{equation}
\cos \theta _{r}=\frac{(1+\beta _{M}^{2})\cos \theta _{0}+2\beta _{M}}{%
1+2\beta _{M}\cos \theta _{0}+\beta _{M}^{2}}.  \label{eq.2}
\end{equation}%
The wave amplitude transforms according to 
\begin{equation}
E_{r}=\mathsf{R}^{1/2}E_{0}\left( \frac{\omega _{r}}{\omega _{0}}\right) ,
\label{eq.3}
\end{equation}%
where $\mathsf{R}$ is the reflection coefficient. The role of the moving
mirror is played by a steep cusps of the electron density in the wake wave,
therefore we write $\beta _{M}=\beta _{ph}=v_{ph}/c$ and $\gamma _{M}=\gamma
_{ph}=1/\sqrt{1-\beta _{ph}^{2}}$. The wave propagates in a narrow angle $%
\Delta \theta \approx 1/\gamma _{M}$. As a result of the interaction with
the moving mirror, the reflected electromagnetic pulse is compressed and its
frequency is upshifted by a factor which in an ultrarelativistic limit, $%
\gamma_{ph}\gg 1$, is approximately equal to $4\gamma_{M}^{2}$. It is also
conceivable that the interaction of the counter-propagating pulse with the
thin relativistic electron shell leads to a higher order harmonic
generation, according to a mechanism considered in details in Refs. \cite%
{OscMir,MTB}. The harmonics allow additional compression with the factor $%
\sim 4n\gamma _{M}^{2}$, where $n$ is a harmonic number.

It is important that the relativistic dependence of the Langmuir frequency
on the wave amplitude results in the formation of wake waves with curved
fronts that have a form close to a paraboloid \cite{D-shape}. The
paraboloidal mirror focuses the reflected pulse, which results in the
intensification of the radiation. In the reference frame moving with the
mirror velocity the reflected light has the wavelength equal to $\approx
\lambda _{0}/2\gamma _{M}$. It can be focused into the spot with the
transverse size $\lambda _{0}/2\gamma _{M}$ with the intensity in the focus
given by 
\begin{equation}
I_{r}=\mathsf{R}I_{0}\gamma _{M}^{6}\left( \frac{D}{\lambda _{0}}\right)
^{2}.  \label{IrIo}
\end{equation}%
Here $I_{0}$ is the incident wave intensity at the pulse waist $D$.

The flying mirror formed by an electron density modulations in the breaking
wake wave can also transform the quasistationary electromagnetic field into
a high-frequency short electromagnetic pulse \cite{RelSol}, such as a
low-frequency electromagnetic field of a relativistic electromagnetic
sub-cycle soliton, quasistatic magnetic field of an electron vortex, and the
quasistationary longitudinal electric field of a wake wave.

The ``flying mirror'' concept was demonstrated in the proof-of-principle
experiments \cite{Kando-2007,Pirozhkov-2007}, where the narrow-band XUV
generation was detected. This opens a way for developing a compact tunable
coherent monochromatic X-ray source with the parameters required for various
applications in biology, medicine, spectroscopy and material sciences\cite%
{Pirozhkov-2007}.

Reasoning from the principal importance of the nonlinear wake wave dynamics
in the ``flying mirror'' concept, in the present paper we consider a variety
of the wake wave breaking regimes. We find typical singularities of the
electron density distribution in the breaking wave and calculate their
reflection coefficients, $\mathsf{R}$. This provides a key for optimising an
operation of the hard electromagnetic radiation source based on the `
`flying mirror'' concept. We describe the EMW interaction with the multiple
cusp structure and show the enhancement of the efficiency of the light
reflection in this configuration.

\section{ Generic properties of the Wake Wave Breaking}

\subsection{Gradient catastrophe}

The wave breaking in collisionless plasmas provides an example of typical
behaviour of the waves in nonlinear systems. The most fundamental properties
of nonlinear waves can be represented by Riemann waves in gas dynamics (e.g.
see Refs. \cite{Z-R,BBK}), described by the equation 
\begin{equation}
\partial _{t}u+w(u)\partial _{x}u=0,  \label{Rie}
\end{equation}%
where $u(x,t)$ is the gas velocity and $w(u)=u+c_{s}(u)$. The sound speed $%
c_{s}(u)$ is a given function of $u$, in a particular case of a cold gas of
noninteracting particles it is equal to zero. The solution to Eq. (\ref{Rie}%
) can be written in an implicit form 
\begin{equation}
u=U_{0}(x-w(u)t)  \label{Rie-sol}
\end{equation}%
with $U_{0}(x)$ being an atrbitrary function deternmined by the initial
condition: $u(x)|_{t=0}=U_{0}(x)$. In a finite time the Riemann wave breaks
at the location where the gradient of the function $u(x,t)$ becomes
infinite. Taking the derivative of the expression (\ref{Rie-sol}) with
respect to $x$, we obtain 
\begin{equation}
\partial _{x}u=\frac{U_{0}^{\prime }(x-w(u))t)}{1+U_{0}^{\prime
}(x-w(u)t)w^{\prime }(u)t}.  \label{grad-u}
\end{equation}%
Here $U_{0}^{\prime }$ and $w^{\prime }$ stand for the derivatives of the
functions $U_{0}$ and $w$ with respect to their arguments. For nontrivial
dependences of $U_{0}(x)$ and $c_{s}(u)$, we find that the denominator in
Eq. (\ref{grad-u}) vanishes at some point $x_{br}$, i. e. the gradient $%
u(x,t)$ tends to infinity, at time $t_{br}$ equal to $t_{br}=-1/U_{0}^{%
\prime }(x_{br}-w(u)t_{br})w^{\prime }(u)$ while the gas velocity $%
u_{br}=U_{0}(x_{br}-w^{\prime }(u_{br})t_{br})$ remains constant. This
phenomenon is known as the ``gradient catastrophe'' or the ``wave breaking''.

Another approach for description of the wave breaking uses a perturbation
theory \cite{BBK} to find the solution to the equation (\ref{Rie}). We write 
\begin{equation}
u=u^{(0)}+\varepsilon u^{(1)}+\varepsilon ^{2}u^{(2)}+\ldots
\end{equation}%
with $\varepsilon \ll 1$. We assume that in zeroth order the wave amplitude
is homogeneous with the velocity, $u^{(0)}$, constant in space and time. To
the first order in the wave amplitude, we have 
\begin{equation}
\partial _{t}u^{(1)}+w(u^{(0)})\partial _{x}u^{(1)}=0.  \label{Rie-u1}
\end{equation}%
This is the simplest wave equation which describes the wave with the
frequency and wavenumber related to each other via the dispersion equation $%
\omega =kw(u^{(0)})$. Thus we obtain the wave propagating in nondispersive
media where both the phase velocity, $\omega /k$, and the group velocity, $%
\partial \omega /\partial k$, are equal to $w(u^{(0)})$. %
The solution to Eq. (\ref{Rie-u1}) is an arbitrary function of the variable $%
x-w_0 t$, where $w_0 = w(u^{(0)})$. We choose it in the form 
\begin{equation}
u^{(1)}(X)=u_{m}\sin k (x-w_0 t).  \label{Rie-u1b}
\end{equation}
To the second order in the wave amplitude, we obtain 
\begin{eqnarray}
\partial_{t}u^{(2)} - w(u^{(0)}) \partial_x u^{(2)}=
-u^{(1)}w^{\prime}(u^{(0)})\partial _{x}u^{(1)}=
\nonumber \\
= -\frac{bk}{2}\sin
\left(2k(x-w_0 t)\right)  \label{Rie-u2}
\end{eqnarray}
with $b=u_{m}^{2}w^{\prime }(u^{(0)})$. The solution to this equation, 
\begin{equation}
u^{(2)}(x,t)=-\frac{bkt}{2}\sin \left(2k(x-w_0 t)\right),  \label{Rie-u2a}
\end{equation}%
describes the second harmonic with the resonant growth of the amplitude in
time.

To the third order in the wave amplitude, we can find that, in general, the
amplitude of the the third harmonic grows with time as $t^2$, and so on. In
the media without dispersion high harmonics are always in resonance with the
first harmonic. The resonance between harmonics appears because of the fact
that the velocity of propagation is the same for all harmonics; it does not
depend on the wave number. This leads to the increase of the velocity
gradient (the wave steepenning) linearly with time and to the break of even
weak but finite amplitude wave.

The situation with the nonlinear Langmuir wave breaking is different. For
simplicity we assume that the wave amplitude is nonrelativistic and the ions
are immobile. We cast the equations of the electron fluid motion and the
electric field induced in the plasma in the form 
\begin{equation}
\partial _{t}v+v\partial _{x}v=-\frac{e}{m_{e}}E,  \label{Lan-v}
\end{equation}%
\begin{equation}
\partial _{t}E+v\partial _{x}E=4\pi en_{0}(\varepsilon x)v.  \label{Lan-E}
\end{equation}%
Here $n_{0}(\varepsilon x)$ with $\varepsilon \ll 1$ is the ion density
which is assumed to be weakly inhomogeneous. Although with the use of the
Lagrange coordinates a solution to Eqs. (\ref{Lan-v},\ref{Lan-E}) can be
reduced to quadratures, we use a perturbation approach in order to analyse
whether or not the high harmonics are in resonance with the first harmonic
in the case of nonlinear Langmuir waves. We expand $v(x,t)$, $E(x,t)$, and $%
n_{0}(\varepsilon x)$ into series: 
\begin{eqnarray}
v &=&\varepsilon v^{(1)}+\varepsilon ^{2}v^{(2)}+\ldots ,  \notag \\
E &=&\varepsilon E^{(1)}+\varepsilon ^{2}E^{(2)}+\ldots \ .
\end{eqnarray}

In first order from Eqs (\ref{Lan-v},\ref{Lan-E}) we obtain the equations 
\begin{equation}
\partial _{t}v^{(1)}+\frac{e}{m_{e}}E^{(1)}=0,
\end{equation}%
\begin{equation}
\partial _{t}E^{(1)}-4\pi en_0(0) v^{(1)}=0
\end{equation}%
with the solution 
\begin{equation}
v^{(1)}=v_{m}\sin (\omega _{pe}t-kx) ,
\end{equation}%
\begin{equation}
E^{(1)}=-\frac{m_{e}v_{m}\omega _{pe}}{e}\cos (\omega _{pe}t-kx) ,
\end{equation}%
where $\omega _{pe}=\sqrt{4\pi n_{0}(0)e^{2}/m_{e}}$ is the Langmuir
frequency.

The second order in the wave amplitude yields%
\begin{equation}
\partial _{t}v^{(2)}+\frac{e}{m_{e}}E^{(2)}=\frac{v_{m}^{2}k}{2}\sin
\left(2(\omega _{pe}t-kx)\right),
\end{equation}%
\begin{eqnarray}
\lefteqn{}&&
\partial _{t}E^{(2)}-4\pi e n_0(0)v^{(2)}=
\nonumber \\
&&
=\frac{m_{e}v_{m}^{2}\omega _{pe}k}{%
2e}\left[ 1-\cos \left(2(\omega _{pe}t-kx)\right) \right]
\nonumber \\
&&
+ 4 \pi e
n_0^\prime (0) x v_m \sin(\omega _{pe}t-kx).
\end{eqnarray}
We obtain for $v^{(2)}$ and $E^{(2)}$%
\begin{eqnarray}
\label{Lan-v-2}
v^{(2)}=-\frac{v_{m}^{2}k}{2\omega _{pe}}\left[ 1+\cos 2(\omega _{pe}t-kx)\right]
+\frac{n_0^\prime(0) x v_m}{4n_0(0)}
\nonumber \\
\times \left[ 2\omega_{pe}t
\cos(\omega _{pe}t-kx)-\sin(\omega _{pe}t-kx) \right] ,
\end{eqnarray}
\begin{eqnarray}
\label{Lan-E-2}
E^{(2)}=-\frac{v_{m}^{2}km_{e}}{2e}\sin \left(2(\omega _{pe}t-kx)\right)
+ \frac{\pi e n_0^\prime(0) x v_m}{\omega_{pe}}
\nonumber \\
\times \left[ 2\omega_{pe}t\sin(%
\omega _{pe}t-kx)-\cos(\omega _{pe}t-kx) \right] .
\end{eqnarray}

As we see, in the homogeneous plasma, $n_0=\mathrm{const}$, there is no a
resonance between the modes in the Langmuir wave, in contrast to the Riemann
waves considered above. This is due to the difference between the group and
the phase velocity of the Langmuir wave: its group velocity in the cold
plasma is equal to zero while the phase velocity is finite and is given by
the relation $\mathrm{v_{ph}}=\omega _{pe}/k$. The Langmuir wave break
occurs if the wave is excited with such a strong amplitude, $v_{0}$, that $%
v_{0}>\mathrm{v_{ph}}$, out of the applicability of the approximation used
in Eqs. (\ref{Lan-v},\ref{Lan-E}). As seen in Eqs. (\ref{Lan-v-2},\ref%
{Lan-E-2}), the breaking occurs also in inhomogeous plasma, where due to the
phase mixing effect the wavenumber, $k$, grows with time \cite{PhMix}.

\subsection{Structure of the Breaking Relativistic Wake Wave}

In the particular application for the light intensification \cite{Light
intensification}, the characteristic features of the electron density
modulations play the key role in the calculating the EMW reflection
coefficient.

\subsubsection{Wake wave breaking in the relativistic electron beam}

\begin{figure}
\includegraphics[scale=1]{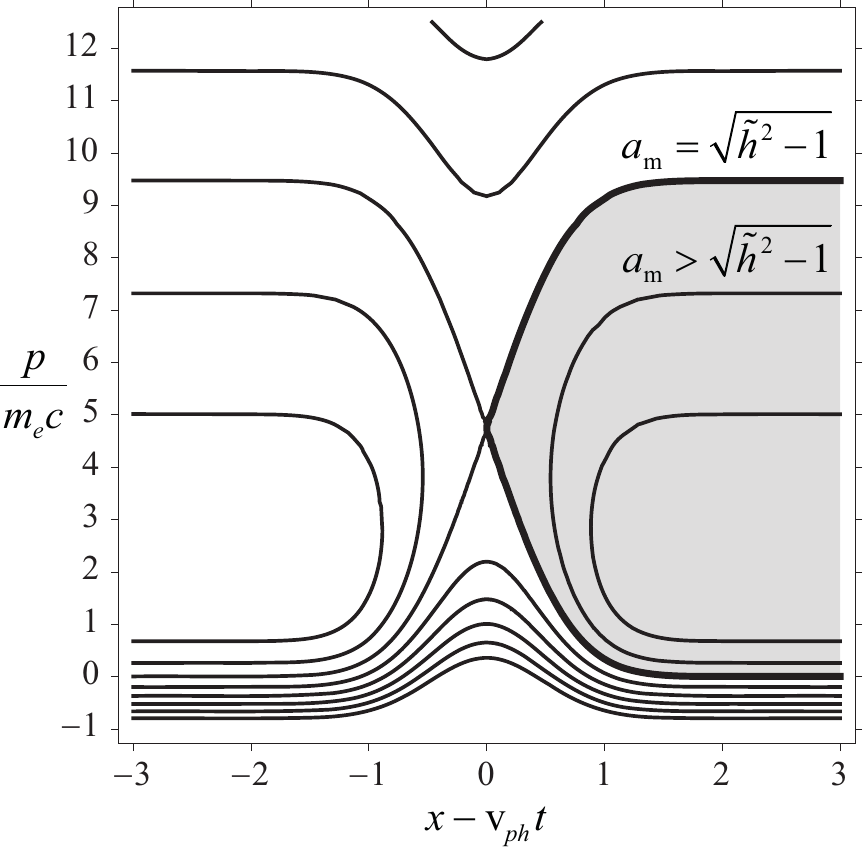}
\caption{Contours of the Hamiltonian Eq. (\ref{el-int}) in
the $(x-{\rm v_{ph}}ct,p)$-plane
for the Gaussian laser pulse $a({\rm X})=a_{\rm m} \exp(-{\rm X}^2)$.
Shaded region represents trajectories of particles
whose initial velocity is in the interval $0\le v\le {\rm v_{ph}}$.
If the trajectory with $h=1$ is the separatrix,
then $a_{\rm m}=\sqrt{\tilde{h}^2-1}$,
while an enclosed trajectory means $a_{\rm m}>\sqrt{\tilde{h}^2-1}$.
}
\label{fig:Ham}
\end{figure}

We start a discussion of the nonlinear wake wave from the consideration of
nonlinear perturbations produced by the laser pulse in the relativistic
electron beam. We neglect the space charge effect assuming that the laser
pulse --- beam interaction can be described in the test particle
approximation. The equation for the electron momentum reads 
\begin{equation}
\partial _{t}p+v\partial _{x}p=-\frac{m_{e}c^{2}}{2\gamma }\partial
_{x}a^{2}.  \label{p-beam}
\end{equation}%
Here the longitudinal component of the electron velocity, $v$, along the
laser pulse propagation direction, is related to the longitudinal component
of the electron momentum by $v=p/\gamma m_{e}$; $\gamma =\sqrt{%
1+a^{2}+\left( {p}/{m_{e}c}\right) ^{2}}$ is the Lorentz-factor and $%
a=eE_{0}/m_{e}\omega _{0}c$ is the normalized electric field of the EMW. We
assume that the EMW is circularly polarized and depends on the coordinate
and time as $a(x-\mathrm{v_{ph}}t)$. The phase velocity of the wake, $%
\mathrm{v_{ph}}$, equals the laser pulse group velocity, \cite%
{Liu-Ros,T-D,Esarey-IEEE}. We note that the EMW group velocity, in general, 
is below the speed of light in vacuum if the laser pulse propagates inside a
waveguide, in a plasma, or/and in the focus region (e.g. see Ref. \cite%
{Esarey}). Introducing a new variable 
\begin{equation}
\mathrm{X}=x-\mathrm{v_{ph}}t,  \label{Xdef}
\end{equation}%
we find the integral of Eq. (\ref{p-beam}) 
\begin{equation}
\gamma -\frac{\mathrm{\beta _{ph}}p}{m_{e}c}=h  \label{el-int}
\end{equation}%
This integral plays a role of the Hamiltonian
in terms of variables ${\rm X}$ and $p$, Fig. \ref{fig:Ham}.
with $h$ being a constant and $\mathrm{\beta _{ph}=v}_{ph}/c$. If the laser
pulse is of a finite duration and the electron momentum is $p_{0}$ before
the interaction with the laser pulse, then the constant $h$ is given by the
expression $h=\sqrt{1+\left( {p}_{0}/{m_{e}c}\right) ^{2}}-\mathrm{\beta
_{ph}}p_{0}/m_{e}c$. When the electron is interacting with the laser pulse
its velocity is given by the expression 
\begin{equation}  \label{v-via-h}
v=c\frac{h\mathrm{\beta _{ph}}-\sqrt{h^{2}- (1-\mathrm{\beta _{ph}^{2}}%
)(1+a^{2}({\mathrm{X}}))}} {h-\mathrm{\beta _{ph}}\sqrt{h^{2}-(1-\mathrm{%
\beta _{ph}^{2}})(1+a^{2}({\mathrm{X}}))}}.
\end{equation}%
The solution of the continuity equation, 
\begin{equation}
\partial _{t}n+\partial _{x}(nv)=0,  \label{cont}
\end{equation}%
gives for the electron density $n=n_{0}(\mathrm{v_{ph}}-v_{0})/(\mathrm{%
v_{ph}}-v)$, which is equivalent to 
\begin{equation}
n=n_{0} \frac{\sqrt{\tilde{h}^{2}-1}}{\tilde{h}-\mathrm{\beta _{ph}}\sqrt{%
\tilde{h}^{2}-1}} \frac{\tilde{h}-\mathrm{\beta _{ph}}\sqrt{\tilde{h}%
^{2}-1-a^{2}({\mathrm{X}})}}{\sqrt{\tilde{h}^{2}-1-a^{2}({\mathrm{X}})}} .
\label{n(a)}
\end{equation}%
Here $v_0=cp_0/(m_e c h+\mathrm{\beta _{ph}}p_{0})$ and 
\begin{equation}
\tilde{h}=\frac{h}{\sqrt{1-\mathrm{\beta _{ph}^{2}}}}.
\end{equation}

At the point $X=X_{br}$, where the denominator in Eq. (\ref{n(a)}) vanishes,
i.e. the electron velocity becomes equal to the phase velocity $\mathrm{%
v_{ph}}$, the electron density tends to infinity. This corresponds to the
breaking of the wave induced by the laser pulse in the electron beam. The
threshold corresponding to the lowest amplitude of the laser pulse, at which
the pulse piles up the particles towars the infinite density, is determined
by the expression 
\begin{equation}
a^{2}({\mathrm{X}}_{br})=\tilde{h}^{2}-1.  \label{a-br}
\end{equation}
In the phase plane of the Hamiltonian system described by
Eq. (\ref{el-int}), the breaking corresponds to $X$-points
and to vertical tangents of the contours of the Hamiltonian,
Fig. \ref{fig:Ham}.

First, we consider the case when the condition (\ref{a-br}) is satisfied
precisely at the maximum of the laser pulse amplitude, $a_{\mathrm{max}}=a({%
\mathrm{X}}_{br})$. In the vicinty of the maximum, the amplitude can be
represented by the expansion 
\begin{equation}
a({\mathrm{X}})=a({\mathrm{X}}_{br})-\frac{1}{2}a^{\prime \prime }({\mathrm{X%
}}_{br})\Delta {\mathrm{X}}^{2},  \label{a0a2}
\end{equation}%
where $\Delta {\mathrm{X=X}}-{\mathrm{X}}_{br}\rightarrow 0$, thus for the
electron density, velocity and momentum we obtain 
\begin{equation}
n=\frac{\bar{n}_{1}}{|\Delta {\mathrm{X}}|},  \label{n(1/x)}
\end{equation}%
\begin{equation}
v=\mathrm{v_{ph}}-\mathrm{\bar{v}_{1}}|\Delta {\mathrm{X}}|,
\label{v-at-xbr}
\end{equation}%
and 
\begin{equation}
p=p_{m}-\bar{p}_{1}|\Delta {\mathrm{X}}|,  \label{p-at-xbr}
\end{equation}%
respectively. Here 
\begin{equation}
\bar{n}_{1}=\frac{n_{0}\tilde{h}a({\mathrm{X}}_{br})}{(\tilde{h}-\mathrm{%
\beta _{ph}}a({\mathrm{X}}_{br}))\sqrt{a({\mathrm{X}}_{br})a^{\prime \prime
}(X_{br})}},
\end{equation}%
$\mathrm{\bar{v}_{1}}={c(1-\mathrm{\beta _{ph}^{2}})\sqrt{a({\mathrm{X}}%
_{br})a^{\prime \prime }(X_{br})}}/{\tilde{h}}$,$\ p_{m}=\mathrm{\beta _{ph}}%
\tilde{h}/\sqrt{1-\mathrm{\beta _{ph}^{2}}}$ and $\bar{p}_{1}=\sqrt{a({%
\mathrm{X}}_{br})a^{\prime \prime }(X_{br})/\mathrm{(1-\mathrm{\beta
_{ph}^{2}})}}$.
\begin{figure}
\includegraphics[scale=1]{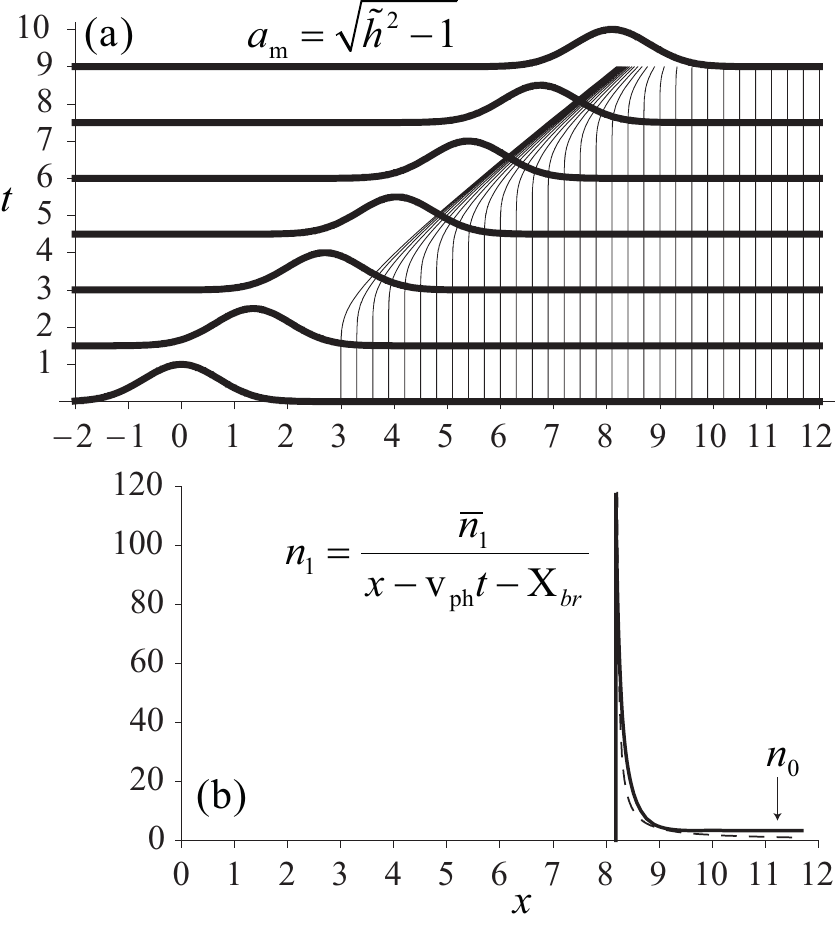}
\caption{Formation of the density singularity
in the beam of particles described by Eq. (\ref{p-beam})
for the Gaussian laser pulse $a({\rm X})=a_{\rm m} \exp(-{\rm X}^2)$,
$a_{\rm m}=\sqrt{\tilde{h}^2-1}$,
and $\beta_{\rm ph}=0.9$.
(a) Thin lines represent the particle trajectories
in the $(x,t)$-plane corresponding to different initial
locations of particles in the beam.
Thick lines stand for the laser pulse envelope
at the corresponding time.
(b) The density of the particles at the last moment of time.
The dashed line shows the theoretical limit Eq. (\ref{n(1/x)}).
}
\label{fig:one}
\end{figure}
The electron motion leading to a pile-up and their density increase
is shown in Fig. \ref{fig:one}.
We note that the singularity in the electron density distribution (\ref{n(1/x)}%
) is not integrable. The integral $\int_{-\Delta {\mathrm{X}}}^{+\Delta {%
\mathrm{X}}}n(X)dX$ diverges. Formally it takes an infinite time in order to
form the singularity. Behavior of the electron density in the vicinity of
the breaking point can be described by using the continuity equation (\ref%
{cont}), the motion integral (\ref{el-int}) and the expansion (\ref{a0a2}).
The electron displacement is described by the equation 
\begin{equation}
\Delta {\mathrm{\dot{X}}} = v-\mathrm{v}_{ph} =-\mathrm{\bar{v}_1} |\Delta 
\mathrm{X}|
\end{equation}%
where a dot stands for the total derivative with respect to time. 
For the initial condition $\Delta {\mathrm{X}}|_{t=0}=\Delta {\mathrm{X}}%
_{0} $
the solution is 
\begin{equation}
\Delta {\mathrm{X}}=\Delta {\mathrm{X}}_{0}\exp (-t/\tau ),
\end{equation}%
where $\tau = {\tilde h}/{(1-\mathrm{\beta_{ph}^2})c\sqrt{a({\mathrm{X}}%
_{br})a^{\prime \prime }(X_{br})}}$ is the singularity pile-up time. The
electron density $n=n_{0}\left \vert \Delta {\mathrm{X}}_{0}/\Delta {\mathrm{%
X}}\right \vert $ at the breaking point grows exponentially: $n=n_{0}\exp
(t/\tau )$. Its growth saturates due to a finite number of the electrons in
the beam or/and due to the space charge effect. The case when the space
charge effect is taken into account is discussed in the next subsection. For
a finite number of the particles, the density distribution asimptotically in
time can be approximated by 
\begin{equation}
n(\mathrm{X})=n_{0}l_{b}\delta ({\mathrm{X}}),
\end{equation}%
where $\delta (x)$ is the Dirac delta function, and $l_{b}$ is the initial
length of the bunch.

Apparently, no break occurs for the phase velocity equal to the speed of
light in vacuum because the electron velocity is always less than $c$. In
this limiting case, $\mathrm{\beta _{ph}=1}$, we have $h=\gamma
_{0}-p_{0}/m_{e}c$. The electron Lorentz gamma factor, $\gamma _{0}$, the
momentum, $p_{0}$, and the velocity, $v_{0}$, before the interaction with
the laser pulse can be written via the constant $h$ as $\gamma
_{0}=(1+h^{2})/2h$, $p_{0}=m_{e}c(1-h^{2})/2h$ and $%
v_{0}=c(1-h^{2})/(1+h^{2})$. Inside the laser pulse the electron acquires $%
\gamma =\gamma _{0}+a^{2}({\mathrm{X}})/2h$, $p=p_{0}+m_{e}c\ a^{2}({\mathrm{%
X}})/2h$, and the electron density becomes equal to 
\begin{equation}
n=n_{0}\left[ 1+\frac{a^{2}({\mathrm{X}})}{1+h^{2}}\right] .
\end{equation}%
We see that for the finite amplitude laser pulse the electron density
remains to be finite.

Now we consider the case when the laser pulse amplitude is larger than the
threshold determined by the condition (\ref{a-br}), $a_{\mathrm{max}}>\tilde{%
h}^{2}-1$. In this case the breaking occurs at the slope of the laser pulse
envelope, where the function $a({\mathrm{X}})$ has the expansion 
\begin{equation}
a({\mathrm{X}})= a({\mathrm{X}}_{br})-a^{\prime }({\mathrm{X}}_{br})\Delta {%
\mathrm{X}}.  \label{a-dx}
\end{equation}%
The local profile of the electron density in the vicinity of the point ${%
\mathrm{X=X}}_{br}$ is described by 
\begin{equation}\label{dens-minus-half}
n(\mathrm{X})=\bar{n}_{1/2}\frac{\chi ({\mathrm{X}})}{\Delta {\mathrm{X}}%
^{1/2}}.
\end{equation}%
Here $\chi (x)$ is the unit-step Heviside function ($\chi (x)=0$ for $x<0$
and $\chi (x)=1$ for $x>0$ ) and%
\begin{equation}
\bar{n}_{1/2}= \frac{n_{0}\tilde h a({\mathrm{X}}_{br})} { (\tilde h - 
\mathrm{\beta _{ph}}a({\mathrm{X}}_{br})) \sqrt{2a({\mathrm{X}}%
_{br})a^{\prime}(X_{br})} } ,
\end{equation}

At the breaking point the electron momentum becomes equal to $p_{1/2}=%
\mathrm{\beta _{ph}}\tilde{h}/\sqrt{1-\mathrm{\beta_{ph}^2}}$. The particle
bounces at this point, starting to move in the same direction as the laser
pulse. Its velocity increases from $\mathrm{v}_{ph}$ at ${\mathrm{X=X}}_{br}$
to 
\begin{equation}
v_{\infty }=c\left( \frac{\tilde{h}\mathrm{\beta _{ph}}+\sqrt{\tilde{h}^{2}-1%
}}{\tilde{h}+\mathrm{\beta _{ph}}\sqrt{\tilde{h}^{2}-1}}\right)
\end{equation}%
for ${\mathrm{X\rightarrow \infty }}$. Asymptotically, the electron density
in the reflected beam tends to 
\begin{equation}
n=n_{0} \frac{\tilde{h}+\mathrm{\beta _{ph}}\sqrt{\tilde{h}^{2}-1}}{\tilde{h}%
-\mathrm{\beta _{ph}}\sqrt{\tilde{h}^{2}-1}}.
\end{equation}

\begin{figure}
\includegraphics[scale=1]{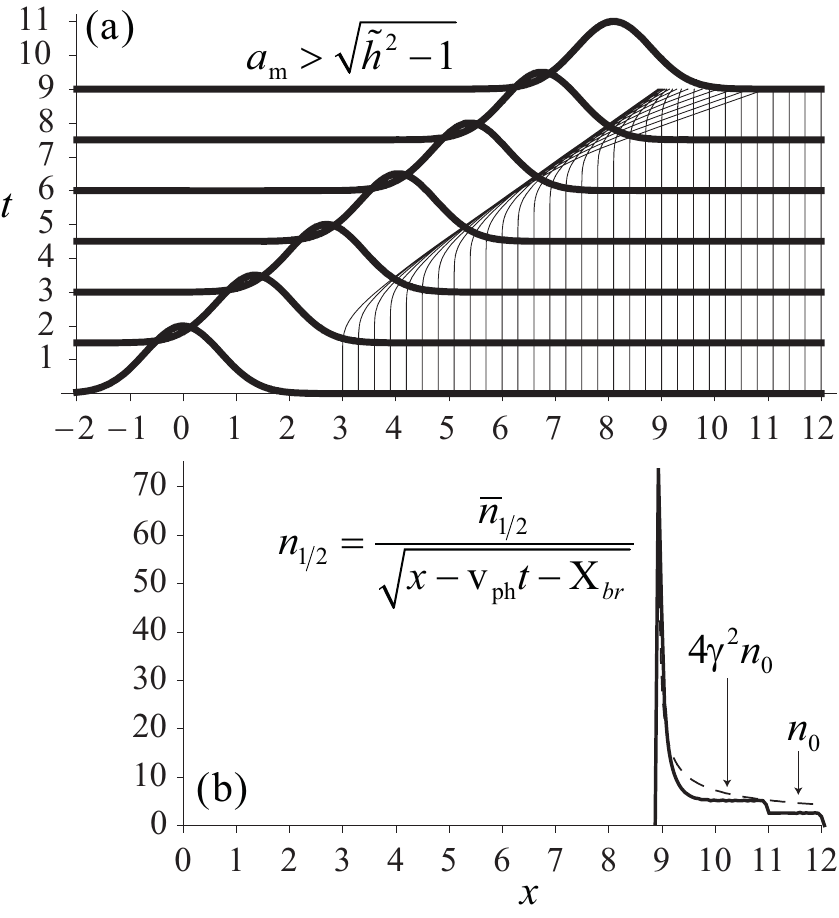}
\caption{The same graphs as in Fig. \ref{fig:one},
but for the two times greater amplitude of the laser pulse,
$a_{\rm m}=2\sqrt{\tilde{h}^2-1}$.
The dashed line in frame (b) shows the main term
in the expansion of the density, Eq. (\ref{dens-minus-half}).
}
\label{fig:half}
\end{figure}

If the particles are at the rest before interacting with the laser pulse,
i.e. $h=1$ and $\tilde{h}=1/\sqrt{1-\mathrm{\beta _{ph}^{2}}}$, then the
density in the reflected beam is equal to $n_{0}(1+\mathrm{\beta _{ph}^{2}}%
)/(1-\mathrm{\beta _{ph}^{2}})$.
The electron motion demonstrating
a bounce of electrons in the ponderomotive potential
of the laser pulse and the corresponding compression of the density,
as well as the electron pile-up at the breaking point,
is shown in Fig. \ref{fig:half}.

According to Eq. (\ref{a-br}) the breaking coordinate depends on the initial
electron momentum. For the electron beam with some energy spread it results
in a non-trivial electron density distribution in the vicinity of the
singularity. We can easily obtain%
\begin{equation}
n(\mathrm{X})=\bar{n}_{1/2}\int \limits_{-\infty }^{+\infty }\frac{\chi ({%
\mathrm{X-X}}_{br}(p_{0}))f(p_{0})dp_{0}}{({\mathrm{X-X}}_{br}(p_{0}))^{1/2}}
\label{n-th}
\end{equation}%
with $f(p_{0})$ being the distribution function normalized on unity: $%
\int_{-\infty }^{+\infty }f(p)dp=1$. We assume that the distribution
function is given by the "water bag" model, i.e. $f(p_{0})=\chi
(p_{0,T}-|p_{0}|)/2p_{0,T}$, with $p_{0,T}/m_{e}c\ll \mathrm{\beta _{ph}}$.
Integration in the right hand-side of Eq. (\ref{n-th}) yields 
\begin{eqnarray}
n(\mathrm{X})=\frac{\bar{n}_{1/2}}{p_{0,T}}
\left[
\chi ({\mathrm{X-X}}_{br}(-p_{0,T}))\sqrt{{\mathrm{X-X}}_{br}(-p_{0})}
\right.
\nonumber \\
\left.
-\chi ({\mathrm{X-X}}_{br}(p_{0,T}))\sqrt{{\mathrm{X-X}}_{br}(p_{0})}
\right].
\end{eqnarray}

The above consideration is carried out within the framework of the test
particle approximation, i.e. neglecting the space charge effect, applicable
when the electron density is low enough. We note that the space charge
effects are negligible in the electron-positron plasmas because the electron
and positron has equal absolute walues of the charge and mass.

Another important case when one can neglect the space charge effects is the
limit of extremely high EMW intensity in the so-called ``radiation pressure
dominated regime'' (for details see Refs. \cite{RadF,TZE,FB}) briefly
considered in Appendix A.

\subsubsection{The plasma wake wave breaking}

Here we discuss the generic properties of the breaking of the Langmuir waves
in plasmas. We shall consider the case when the wake wave is excited by a
high intensity laser pulse \cite{Liu-Ros,T-D,GK}. This consideration is
relevant to the breaking of the wake wave excited by the ultrarelativistic
electron bunch \cite{Chen}. Within the framework of one-dimensional
approximation, the relativistic Langmuir wave excitation by the
electromagnetic wave can be described by the equation for the electron
momentum, $p$, 
\begin{equation}
\partial _{t}p+v\partial _{x}p=-eE-\frac{m_{e}c^{2}}{2\gamma }\partial
_{x}a^{2}.  \label{p1}
\end{equation}%
In contrast to Eq. (\ref{p-beam}), it incorporates the space charge effects
via the term with the electric field, $E$, for which we have the equation 
\begin{equation}
\partial _{t}E+v\partial _{x}E=4\pi en_{0}ev.  \label{E2}
\end{equation}%
Left in a plasma behind a finite width laser pulse, the Langmuir wave has
the period $T_{wf}={2\pi }/{\omega _{pe}}=\sqrt{{m_{e}}/{2ne^{2}}}$, in the
limit of small amplitude, ${p_{m}}/{m_{e}c}\rightarrow 0$. In the opposite
limit of a large amplitude, ${p_{m}}/{m_{e}c}\gg 1$, the wake wave period is 
$T_{wf}=4\sqrt{2p_{m}/m_{e}c}/{\omega _{pe}}$. Here the maximum wave
amplitude, $p_{m}$, is related to the laser pulse amplitude as $p_{m}=m_{e}c{%
a^{2}}/2$. Therefore, the wakefield wavelength $\lambda _{wf}\approx
cT\propto a$ is proportional to the laser pulse amplitude.

For the laser pulse propagating with constant phase velocity $\mathrm{v_{ph}}
$, we seek a solution of Eqs. (\ref{p1},\ref{E2}) in the form of a
progressive wave, when the solution depends on the variables $t$ and $x$ in
the combination (\ref{Xdef}). As a result we obtain the ordinary
differential equation 
\begin{equation}
\left( \gamma -\frac{p}{m_{e}c^{2}}\mathrm{v_{ph}}\right) ^{\prime \prime }=%
\frac{\omega _{pe}^{2}}{c^{2}}\frac{p}{\gamma m_{e}\mathrm{v_{ph}}-p},
\label{ddp}
\end{equation}%
where the prime denotes a derivative with respect to $\mathrm{X}$. For
piecewise-constant profiles of $a({\mathrm{X}})$, a solution of Eq. (\ref%
{ddp}) can be expressed in terms of elliptic integrals. Here we analyze the
solutions of this equation in the vicinity of the singularity. The right
hand side of Eq. (\ref{ddp}) becomes singular when the denominator, $\gamma
m_{e}\mathrm{v_{ph}}-p$, tends to zero, i.e when the electron velocity $v$
becomes equal to the pase velocity of the wake wave, $\mathrm{v_{ph}}$. In
the stationary wake wave, the singularity is reached at the maximum value of
the electron velocity, $v_{m}={p_{m}}/{\gamma _{m}m_{e}}$, at $\mathrm{X}=%
\mathrm{X}_m$. This singularity, called also the ``gradient catastrophe'',
corresponds to the wave breaking. In other words, the Langmuir wave breaks
when the electric field $E_{wf}$ is above the Akhiezer-Polovin electric
field $E_{A-P}=(m_{e}\omega _{pe}c/e)\sqrt{2(\gamma _{ph}-1)}$ with $\gamma
_{ph}=1/\sqrt{1-\mathrm{\beta _{ph}^{2}}}$, which means that the electron
displacement inside the wave becomes equal to or larger than the wavelength
of the wake plasma wave (see Ref. \cite{Akhiezer-Polovin}).

In order to find the singularity structure we expand the electron momentum, $%
p$, and the Lorentz factor, $\gamma$, in the vicinity of their maximum, $%
\Delta \mathrm{X}=\mathrm{X}-\mathrm{X}_m \rightarrow 0$. The momentum is
represented by 
\begin{equation}
p=p_{m}+\delta p + O(\delta p^2),
\end{equation}%
with 
\begin{equation}
p_{m}=m_{e}c\mathrm{\beta _{ph}}\sqrt{\frac{1+a_m^{2}}{1-\mathrm{\beta
_{ph}^2}}}
\end{equation}%
and $|\delta p/p_{m}|\ll 1$. Here $a_m = a({\mathrm{X}_m})$. Keeping the
main terms of expansions over $\delta p$ in both sides of Eq. (\ref{ddp}),
we obtain 
\begin{equation}
\left( \delta p^{2}\right) ^{\prime \prime }=-\varkappa ^{2}\frac{1}{\delta p%
}  \label{delp}
\end{equation}%
with $\varkappa ^{2}= 2 \mathrm{\beta_{ph}}\gamma_{\mathrm{ph}}^6
(1+a_m^{2})\omega_{pe}^{2} m_{e}^3 c$. Multiplying the left- and right-hand
sides of Eq. (\ref{delp}) on $(\delta p^2)^\prime$ and integrating over ${%
\mathrm{X}}$, we obtain 
\begin{equation}
\delta p\delta p^{\prime }=\sqrt{m_e^2 c^2 f^2-\varkappa ^{2}\delta p},
\label{dpdp}
\end{equation}%
where $f$ is an integration constant. We note that since Eq. (\ref{delp})
has singularity due to the initial condition $\left.\delta p\right|_{\Delta 
\mathrm{X}=0}=0$, in principle, one can choose different values for the
integration constant in different intervals with boundary $\Delta \mathrm{X}%
=0$, i.~e. $f=f^-$ in the interval $\Delta \mathrm{X}<0$ and $f=f^+ \not=
f^- $ for $\Delta \mathrm{X}>0$. The behavior of the function $\delta p({%
\mathrm{X}})$ at the singularity is different depending on the behaviour of
the product $\delta p\delta p^{\prime }$ at $\delta p\rightarrow 0$.

If the product $\delta p\delta p^{\prime }$ is finite and nonzero for $%
\delta p\rightarrow 0$, then the main term in the expansion of the solution
of Eq. (\ref{dpdp}) at $\Delta \mathrm{X}\rightarrow 0$ is 
\begin{equation}
\delta p=-m_e c \sqrt{2f|\Delta {\mathrm{X}}|}.
\end{equation}%
Using this expression we find the electron velocity: 
\begin{equation}
v\approx \mathrm{v_{ph}}- \frac{c}{\gamma_{ph}^3} \left( \frac{2f|\Delta {%
\mathrm{X}}|}{1+a_m^2} \right)^{1/2}
\end{equation}%
The electron density, $n(\mathrm{X})$, is given by the solution to the
continuity equation (\ref{cont}). In the vicinity of the singularity it
reads 
\begin{equation}
n=\frac{n_{0}\mathrm{v_{ph}}}{\mathrm{v_{ph}}-v} \approx n_0 \mathrm{%
\beta_{ph}} \gamma_{ph}^3 \left( \frac{1+a_m^2}{2f|\Delta {\mathrm{X}}|}
\right)^{1/2}.  \label{nx1/2}
\end{equation}

If at $\delta p\rightarrow 0$, $\delta p\delta p^{\prime }\rightarrow 0$,
which means $f=0,$ then the main term in the expansion of the solution of
Eq. (\ref{dpdp}) at $\Delta \mathrm{X}\rightarrow 0$ reads 
\begin{eqnarray}
\lefteqn{}&&
\delta p= -\left(\frac{3}{2} \varkappa \Delta \mathrm{X}\right)^{2/3} =
\nonumber \\
&&
=
-m_e 
\mathrm{\beta_{ph}}\gamma_{ph}^3 \left( \frac{3\omega_{pe}\sqrt{1+a_m^2}}{%
\sqrt{2}\mathrm{\beta_{ph}}} \Delta {\mathrm{X}} \right)^{2/3}
\end{eqnarray}
For the electron velocity we obtain 
\begin{equation}
v\approx \mathrm{v_{ph}} - \frac{\mathrm{v_{ph}}}{\gamma_{ph}} \left( \frac{%
3\omega_{pe}(1+a_m^2)^{1/4}}{\sqrt{2}\mathrm{\beta_{ph}}} \Delta {\mathrm{X}}
\right)^{2/3}  \label{v2/3}
\end{equation}%
In the vicinity of the singularity the electron density distribution is
given by 
\begin{equation}
n=\frac{n_{0}\mathrm{v_{ph}}}{\mathrm{v_{ph}}-v} \approx n_0\gamma_{ph}
\left( \frac{\sqrt{2}(1+a_m^2)^{1/4} \mathrm{\beta_{ph}} } {3\omega_{pe}
\Delta {\mathrm{X}}} \right)^{2/3}  \label{n2/3}
\end{equation}

We found three types of the singularity which appears in the electron
density due to wave breaking. In each case the singularity corresponds to
the caustic in the plasma flow with inhomogeneous velocity. General
description evoking geometric properties of the hydrodynamic flow, presented
in Appendix B, shows that the electron density singularity can be
approximated by the dependence $n({\mathrm{X}})\propto ({\mathrm{X}-\mathrm{X%
}_m})^{-\alpha}$ with $1/2 \le \alpha <1$.

\section{Electromagnetic wave reflection at the crest of the breaking wake}

As we discussed in Introduction, the ``flying mirror'' concept proposed in
Ref. \cite{Light intensification} makes use of the wave breaking occuring in
wake waves generated in a subcritical plasma by an ultrashort laser pulse
(further called ``driver''). In a strongly nonlinear wake wave, the electron
density is modulated in such a way that the electrons form relatively thin
layers moving with the velocity $v_{ph}$. A counter-propagating laser pulse
(called ``signal'') interacts with the plasma, perturbed by the driver.
Under certain conditions, the signal is partially reflected from the wake
wave, which thereby plays the role of a relativistic mirror. As the wake
wave amplitude approaches the threshold for wave breaking (i.e., when the
electron velocity in the wave approaches its wave phase velocity), the
electric field profile in the wave steepens and, correspondingly, the
singularity is formed in the electron density profile. It is important that
the singularity affords partial reflection of the signal with some
reflection coefficient $\mathsf{R}$. As shown in Appendix B, in general, the
singularity can be represented as $n({\mathrm{X}})\propto ({\mathrm{X}-%
\mathrm{X}_m})^{-\alpha}$, $1/2 \le \alpha <1$. Although the singularity is
integrable for $\alpha<1$, it breaks the geometric optics approximation and
leads to the efficient reflection of a portion of the radiation accompanied
by the upshifting of the frequency of the reflected pulse. When $\alpha
\rightarrow 1$, the modulation of the electron density is so strong that the
reflection coefficient becomes to be of the order of unity.

In order to calculate the reflection coefficient, we consider the
interaction of an electromagnetic wave, representing the signal, with a
singularity of the electron density formed in a breaking Langmuir wave. The
electromagnetic wave, given by the $z$-component of the vector potential $%
A_z (x,y,t)$, is described by the wave equation 
\begin{equation}
\partial _{tt}A_{z}-c^{2}(\partial _{xx}A_{z}+\partial _{yy}A_{z})+\omega
_{pe}^{2}(x-\mathrm{v_{ph}}t)A_{z}=0,  \label{wave-eq}
\end{equation}%
where $\omega _{pe}^{2}({\mathrm{X}})=4\pi e^{2}n({\mathrm{X}})/m_{e}\gamma $
and the electron Lorentz factor, $\gamma $, is equal to $\gamma _{ph}$ at
the point where the density, $n(X)$, is maximum.

\subsection{The case $n({\mathrm{X}})\propto {\mathrm{X}}^{-\protect \alpha }$
with $\protect \alpha <1$}

Here we consider the general case of the singularity formed in the electron
density at the point of the wake wave breaking, which is defined by 
\begin{equation}
n({\mathrm{X}}) = \frac{n_0 G_\alpha}{(k_p\mathrm{X})^{\alpha}},
\end{equation}
where $G_\alpha = \mathrm{const}$ is the dimensionless constant, $1/2\leq
\alpha < 1 $.

In the boosted reference frame, moving with the phase velocity of the
Langmuir wave, the equation (\ref{wave-eq}) takes the form 
\begin{equation}
\frac{d^{2}a(\zeta )}{d\zeta ^{2}}+\left( s^{2}-\frac{g_{\alpha }}{|\zeta
|^{\alpha }}\right) a(\zeta )=0  \label{eqxalpha}
\end{equation}%
with 
\begin{equation}
a(\zeta )=\frac{eA_{z}(\zeta )}{m_{e}c^{2}}\exp (-i({\omega }^{\prime }{t}%
^{\prime }-{k}_{y}{y}))
\end{equation}%
and 
\begin{equation}
g_{\alpha }=G_{\alpha }k_{p}^{2-\alpha }\gamma _{ph}^{\alpha -1}.
\label{galpha}
\end{equation}%
Here 
\begin{equation}
s^{2}=(\omega ^{\prime }/c)^{2}-k_{y}^{2}>0,
\end{equation}%
and $\zeta =(x-\mathrm{v_{ph}}t)\gamma _{\mathrm{ph}}$, ${t}^{\prime }$, ${k}%
^{\prime },{\omega }^{\prime }$ are the coordinates and time and the wave
vector and frequency in the boosted frame. According to Eqs. (\ref{n2/3}), (%
\ref{galpha}), the coefficient $g_{\alpha }$ for $\alpha =2/3$ is equal to 
\begin{equation}
g_{2/3}\approx ({2}/{9})^{1/3}(1+a_{m}^{2})^{1/6}k_{p}^{4/3}\gamma
_{ph}^{2/3},
\end{equation}%
i.e. $G_{2/3}=({2}/{9})^{1/3}(1+a_{m}^{2})^{1/6}\gamma _{ph}$.

WKB-approximation gives asymptotic solutions of Eq. (\ref{eqxalpha}) for
large $|s|$ and $|\zeta |$ 
\begin{eqnarray}
a(\zeta ) &\approx &W_{1}(\zeta )\left \{ \tau
\,e^{isW(\zeta )}
+\rho 
\,e^{-isW(\zeta )}
\right \} ,  \label{WKB}
\end{eqnarray}
\begin{eqnarray}
W(\zeta ) &=&\int \limits_{\zeta _{0}}^{\zeta }\sqrt{1-\frac{g_{\alpha }}{%
s^{2}|\zeta |^{\alpha }}}d\zeta , \\
W_{1}(\zeta ) &=&\left( 1-\frac{g_{\alpha }}{s^{2}|\zeta |^{\alpha }}\right)
^{-1/4},
\end{eqnarray}
where $\tau $ and $\rho $ are constant.

We are interested in solutions such that for $\zeta \rightarrow +\infty $, $%
\tau $ represents the amplitude of the incident wave (assumed to be equal to
1) and $\rho $ is the reflected wave amplitude, while in the opposite limit, 
$\zeta \rightarrow -\infty $, $\tau $ stands for the transmitted wave and $%
\rho $ vanishes. Consequently, we write: 
\begin{equation}
\tau (+\infty )=1,\quad |\rho (+\infty )|^2=\mathsf{R},  \label{coeff-def1}
\end{equation}%
\begin{equation}
|\tau (-\infty )|^2=\mathsf{T},\quad \rho (-\infty )=0,  \label{coeff-def2}
\end{equation}%
where we assume that the reflection is small, 
\begin{equation}
\mathsf{R}\rightarrow 0\  \quad \mathrm{and}\quad \mathsf{T}\rightarrow
1\quad \mathrm{at}\quad s\rightarrow \infty .
\end{equation}

We multiply Eq. (\ref{eqxalpha}) by $da(\zeta )/d\zeta $, integrate it over $%
\zeta $ and obtain 
\begin{equation}
\left. \left[ \left( \frac{da(\zeta )}{d\zeta }\right) ^{2}+s^{2}a^{2}(\zeta
)\right] \right \vert _{\zeta _{1}}^{\zeta _{2}}=\int \limits_{\zeta
_{1}}^{\zeta _{2}}\frac{g_{\alpha }}{|\zeta |^{\alpha }}\frac{da^{2}(\zeta )%
}{d\zeta }d\zeta \,.
\end{equation}

According to Eq. (\ref{WKB}), at large $|s|$ and $|\zeta |$ we have 
\begin{eqnarray}
\lefteqn{}&&
\left( \frac{da(\zeta )}{d\zeta }\right) ^{2}+s^{2}a^{2}(\zeta )\approx
4s^{2}\tau \rho
\nonumber\\
&&
+\frac{g_{\alpha }}{|\zeta |^{\alpha }}
\left[ \tau ^{2} 
e^{2isW(\zeta )}
+\rho ^{2}
e^{-2isW(\zeta )}
\right] .
\end{eqnarray}
Using Eq. (\ref{coeff-def1}), (\ref{coeff-def2}), we obtain 
\begin{eqnarray}
\mathsf{R}_{\alpha }=\frac{1}{4s^{2}}\left. \left[ \left( \frac{da(\zeta )}{%
d\zeta }\right) ^{2}+s^{2}a^{2}(\zeta )\right] \right \vert _{-\infty
}^{\infty }=
\nonumber\\
\frac{g_{\alpha }}{4s^{2}}\int \limits_{-\infty }^{+\infty }%
\frac{1}{|\zeta |^{\alpha }}\frac{da^{2}(\zeta )}{d\zeta }d\zeta .
\label{eqR}
\end{eqnarray}
Where the main term in the integrand in the right hand side of Eq. (\ref{eqR}%
) for large $s$ is 
\begin{eqnarray}
\frac{1}{|\zeta |^{\alpha }}\frac{da^{2}(\zeta )}{d\zeta }\approx \frac{%
2isW_{1}^{2}(\zeta )\exp (2isW(\zeta ))}{|\zeta |^{\alpha }}\approx
\nonumber\\
\approx \frac{2is}{|\zeta |^{\alpha }}\exp \left( 2is\zeta -\frac{ig_{\alpha }\zeta
^{1-\alpha }}{(1-\alpha )s}\right) .  \label{eqRHS}
\end{eqnarray}
This approximation fails in an interval around $\zeta =0$. The interval size
decreases at large $s$ and small $g_{\alpha }$ thus the intergration over
this interval yields negligibly small contribution to the reflection
coefficient. Neglecting $s^{-1}$-term in the exponent in Eq. (\ref{eqRHS}),
we obtain 
\begin{equation}
\mathsf{R}_{\alpha }=\frac{ig_{\alpha }\Gamma (1-\alpha )}{2^{1-\alpha
}s^{2-\alpha }}\sin \left( \frac{\pi \alpha }{2}\right) ,  \label{Ralpha}
\end{equation}%
where $\Gamma$ is the Gamma function \cite{A-S}. For $\alpha =1/2$, this
coefficient is 
\begin{equation}
\mathsf{R}_{1/2}=\frac{i\pi^{1/2}g_{1/2}}{2 s^{3/2}}.
\end{equation}
For $\alpha =2/3$, this coefficient is 
\begin{equation}
\mathsf{R}_{2/3}=\frac{i3^{1/2}\Gamma (1/3)g_{2/3}}{(2s)^{4/3}}.
\end{equation}

For $\alpha \rightarrow 1$ and fixed $s$ and $g_{\alpha}$, the main term in
the expansion of the coefficient (\ref{Ralpha}) over $(1-\alpha)$ is $%
\mathsf{R}_{\alpha }\approx ig_{\alpha }s^{-1}(1-\alpha )^{-1}$. The above
used approximation implies a smallness of the relection coefficient compared
to unity, i.e. the condition $s\gg (1-\alpha )^{-1}$ must be fulfilled. In
the case $\alpha =1$, which requires special consideration, this condition
can not be satisfied. The case $\alpha =1$ is analysed below.

\subsection{The case $n({\mathrm{X}})\propto {\mathrm{X}}^{-1}$}

Here we consider the problem of EMW reflection at the following electron
density profile 
\begin{equation}
n({\mathrm{X}})=\frac{n_{0}G_{1}}{k_p|\mathrm{X}|} .  \label{n(X1)}
\end{equation}%
Since this singularity is non-integrable, the thin layer approximation is
not applicable. As stated above, this profile formally requires an infinite
number of electrons and its formation takes an infinite time. We use the
standard matching technique. Substituting $\alpha=1$ to Eq. (\ref{eqxalpha})
we obtain the equation 
\begin{equation}
\frac{d^{2}a(\zeta )}{d\zeta ^{2}}+ \left( s^{2}-\frac{g_{1}}{|\zeta |}%
\right) a(\zeta )=0
\end{equation}
where $\zeta $ is the coordinate in the boosted frame of reference and $g_1$
is defined by Eqs. (\ref{galpha}. The solution to this equation can be
expressed in terms of the confluent hypergeometric functions of the first, $%
M(a,b,z)$, and the second kind, $U(a,b,z)$, also known as the Kummer's
function of the first and second kind \cite{A-S}: 
\begin{eqnarray}
a_{+}(\zeta )&=&i
e^{-is\zeta}
\left[ \frac{-c_{1}^{+}}{g_1\Gamma \left( {ig}%
_{1}/{2s}\right) }U\left( -\frac{ig_{1}}{2s},0,2is\zeta \right)
\right.
\nonumber\\
&&
\left.
+c_{2}^{+}\zeta M\left( 1-\frac{ig_{1}}{2s},2,2is\zeta \right) \right]
\label{a+}
\end{eqnarray}
in the region $\zeta >0$ and 
\begin{eqnarray}
a_{-}(\zeta )&=&i
e^{-is\zeta}
\left[ \frac{c_{1}^{-}}{g_1\Gamma \left( {-ig}%
_{1}/{2s}\right) }U\left( \frac{ig_{1}}{2s},0,2is\zeta \right)
\right.
\nonumber\\
&&
\left.
+c_{2}^{-}\zeta M\left( 1+\frac{ig_{1}}{2s},2,2is\zeta \right) \right]
\label{a-}
\end{eqnarray}
for the negative coordinate $\zeta <0$. These functions should be equal to
each other at $\zeta =0$ for all $s$, i.e. 
\begin{equation}
a_{+}(+0)-a_{-}(-0)=\frac{(c_{1}^{+}-c_{1}^{-})}{\pi g_1}\sinh \left( \frac{%
\pi g_{1}}{2s}\right) =0,
\end{equation}%
which entails $c_{1}^{-}=c_{1}^{+}$. The derivatives of $a_{+}(\zeta )$ and $%
a_{-}(t)$ with respect to $\zeta $ should differ at $\zeta=0$ for all $s$ by 
\begin{eqnarray}
a_{+}^{\prime }(+0)-a_{-}^{\prime }(-0)&=&V.P.\int \limits_{-\varepsilon
}^{+\varepsilon }\frac{a(\zeta )}{|\zeta |}\,d\zeta =
\nonumber\\
&=&
\frac{2c_{1}^{+}}{\pi }%
\ln (\varepsilon )\sinh \left( \frac{\pi g_{1}}{2s}\right) .
\end{eqnarray}
It yields 
\begin{equation*}
c_{2}^{-}=c_{2}^{+}-c_{1}^{+}\sinh \left( \frac{\pi g_{1}}{2s}\right) \left[%
1+i d(s)\right] ,
\end{equation*}%
where 
\begin{eqnarray}
d(s)=&&\frac{1}{\pi} \Big[ 4\mathbf{\gamma}_{\mathrm{E-M}}+2\log \left( {2s} \right)
\Big.
\nonumber\\
&&\Big.
+ \psi \left( 1-\frac{ig_{1}}{2s}\right) +\psi \left( 1+\frac{ig_{1}%
}{2s}\right) \Big] ,
\end{eqnarray}
is real monotonously growing function for $g_1>0$ and $s>0$, $\mathbf{\gamma}%
_{\mathrm{E-M}}=0.577...$ is the Euler-Mascheroni constant and $\psi (z)$ is
the digamma function \cite{A-S}.

The series expansions of both the branches, (\ref{a+}) and (\ref{a-}), at $%
\zeta \rightarrow \pm \infty $ contain the terms proportional to $\exp {(i
s\zeta )}$ and $\exp {(-i s\zeta )}$. The positive exponent corresponds to
the incident and transmitted wave, while the negative exponent -- to the
reflected wave. Obviously there is no reflected wave behind the singularity
at $\zeta \rightarrow -\infty $, i.e. the term in $a_{-}(\zeta)$
proportional to $\exp {(-i s\zeta )}$ should vanish. As a result we find a
relationship between $c_{2}^{+}$ and $c_{1}^{+}$: 
\begin{equation}
c_{2}^{+}=c_{1}^{+}\! \left[ \cosh \left( \frac{\pi g_{1}}{2s}\right) +
i\sinh \left( \frac{\pi g_{1}}{2s}\right) d(s)\right] .
\end{equation}%
Now we can find the amplitudes of the reflected, $\rho $, and transmitted, $%
\tau$, waves as the ratios of the reflected-to-incident and
transmitted-to-incident waves, respectively: 
\begin{eqnarray}
\lefteqn{}&&
\rho (s)=\frac{\Gamma (-ig_{1}/2s)}{\Gamma (ig_{1}/2s)} \left( {2s}\right)^{i g_1/s}
\nonumber\\
&&\times
\left[ 1-\frac{\exp(-\pi g_1/2s)} {\cosh (\pi g_{1}/2s)+i\sinh (\pi g_{1}/2s) d(s)}\right] ,
\\
\lefteqn{}&&
\tau (s)=-\frac{\Gamma (-ig_{1}/2s)}{\Gamma (ig_{1}/2s)} \left({2s}\right)^{i g_1/s}
\nonumber\\
&&\times
\frac{\exp(-\pi g_1/s)} {\cosh (\pi g_{1}/2s)+i\sinh (\pi g_{1}/2s) d(s)} .
\end{eqnarray}
The reflection coefficient is 
\begin{equation}
\mathsf{R}_1 = |\rho|^2= 1- \frac{1}{\cosh^2(\pi g_1/2s) + d(s)^2\sinh^2(\pi
g_1/2s)} ,
\end{equation}
the transmission coefficient is 
\begin{equation}
\mathsf{T}_1 = |\tau|^2= \frac{\exp(-2\pi g_1/s)}{\cosh^2(\pi g_1/2s) +
d(s)^2\sinh^2(\pi g_1/2s)} ,
\end{equation}

\begin{figure}
\includegraphics[scale=1]{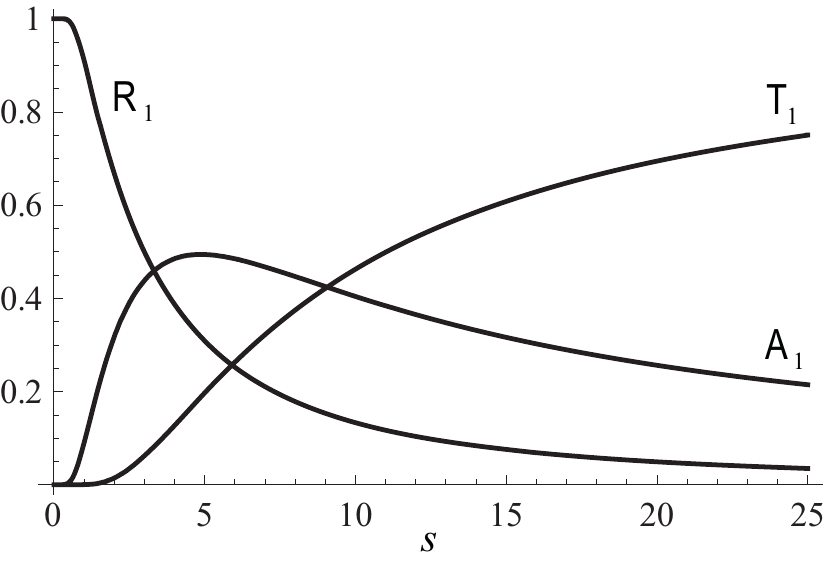}
\caption{Coefficients of reflection ($\mathsf{R}_1=\protect \rho^{2}$),
transmission ($\mathsf{T}_1=\protect \tau^{2}$), and absorbtion ($\mathsf{A}%
_1=1-\mathsf{R}_1-\mathsf{T}_1$) as functions of the wave number $s$
for the case $n(\mathrm{X})\propto \mathrm{X}^{-1}$ .}
\label{fig:coeff}
\end{figure}

We note that for the density profile $n\propto |{\zeta }|^{-1}$ the sum of
the reflection and transmission coefficients is less than one, $\mathsf{R}%
_{1}+\mathsf{T}_{1}=|\rho |^{2}+|\tau |^{2}<1$. This corresponds to a
partial absorbtion of the wave energy in the region of the singularity,
similarly to the case $n\propto {\zeta }^{-1}$ anylized in Refs. \cite%
{Budden} and \cite{Stix}. The dependence of the reflection, transmission and
absorption, $\mathsf{A}_{1}=1-\mathsf{R}_{1}-\mathsf{T}_{1}$, coefficients
on the wave number, $s$, is shown on Fig. \ref{fig:coeff}.

\section{Laser pulse interaction with multiple relativistic mirrors}

A number of the photons reflected at the semi-transparent mirror depends on
the regime of nonlinear wake wave breaking. In Refs. \cite%
{KrSoob,Pirozhkov-2007,BEKT,MTB} and above it is shown that although the
reflection coefficient is small, it is not exponentially smal. For example,
in the case of typical singularity with $n({\mathrm{X}})\propto {\mathrm{X}}%
^{-2/3}$ the reflection coefficient in the boosted frame of reference is $%
\mathsf{R}\approx 1/2\gamma _{M}^{3}$, \textit{i.e.} a number of photons
reflected at a single breaking wake is $\delta N_{ph}\approx
N_{s,ph}/2\gamma _{M}^{3}$. Here $N_{s,ph}$ is a number of photons in the
incident (signal) pulse.

There is a way to increase the number of reflected photons by using multiple
wake wave periods with the electron density singularity in each period. Such
the configuration is typically formed behind the ultrashort laser pulse in
the underdense plasma. It is easy to show that when the pulse interacts with
a train of wake waves, the number of reflected photons is given by the sum 
\begin{equation}
N_{r,ph}=N_{s,ph}r\sum_{i=0}^{\mathcal{N}_{w}} (1-\mathsf{R})^{i}=N_{s,ph}%
\left[ 1-(1-\mathsf{R})^{\mathcal{N}_{w}}\right] .  \label{eq.5-train}
\end{equation}%
Here $\mathcal{N}_{w}$ is a number of the wake wave periods. Then the energy
of the reflected electromagnetic radiation is 
\begin{equation}
\mathcal{E}_{r}=\mathcal{E}_{s}\left( \frac{1+\beta _{M}}{1-\beta _{M}}%
\right) \left[ 1-(1-\mathsf{R})^{\mathcal{N}_{w}}\right] .  \label{eq.6}
\end{equation}%
In the limit $\mathsf{R}\mathcal{N}_{w}\ll 1$, we have $N_{r,ph}\approx
N_{s,ph}\mathsf{R}\mathcal{N}_{w}$, with the reflected electromagnetic
energy given by 
\begin{equation}
\mathcal{E}_{r}\approx (4\mathsf{R}\mathcal{N}_{w}/\gamma _{M}) \mathcal{E}%
_{s} .  \label{eq.7}
\end{equation}%
In the limit $\mathcal{N}_{w}\rightarrow \infty $, the reflected back energy
equals to 
\begin{equation}
\mathcal{E}_{r}=4\gamma _{M}^{2}\mathcal{E}_{s}.  \label{eq.8}
\end{equation}
We see that the energy of the reflected pulse can be many times greater than
that of the incident pulse. The energy gain is due to the momentum transfer
from the wake wave to the reflected radiation.

Under conditions which are expected to be realized in a moderate intensity
laser --- gas target interaction, the plasma inhomogeneity can play an
important role. If the driver pulse normalized amplitude, $a=eE/m_{e}\omega
c $, is not large enough to excite the breaking wake wave, \textit{i.e.} $%
a^{2}<2\gamma _{ph}$, the wave reaches the breaking threshold due to growth
of its wave number. As well known, the Langmuir waves belong to a class of
the waves with continuous frequiency spectrum. In inhomogeneous medium the
wave vector, $\mathbf{k}_{p}$ , and wave frequency, $\omega _{p}$, are
related to each other via the equation 
\begin{equation}
\partial _{t}\mathbf{k}_{p}=-\nabla \omega _{p}.
\end{equation}%
The frequency dependence on the coordinates can also be due its dependence
on the wave amplitude in the case of transversally inhomogeneous laser pulse
driver, which is typical for the laser plasma interaction. As a result, the
wake wave breaks when $\omega _{p}/|\mathbf{k}_{p}|\approx v_{e}$, i.e. when
the wave phase velocity, $\omega _{p}/|\mathbf{k}_{p}|$, becomes equal to
the electron quiver velocity $v_{e}$ (for details see Refs. \cite{PhMix}).
In this regime the breaking wake wave gamma factor, $\gamma _{ph}$, is
determined not by the plasma density, but by the driver laser pulse
amplitude, and it is equal to $\gamma _{e}$.

\section{Conclusion}

The plasma wave breaking leads to the formation of moving caustics in the
plasma flow, which correspond to singularities of the particles distribution
in the phase space. The singularity structure depends of the breaking
conditions determined by the parameters of plasma and the driver laser pulse
which excites the plasma wave. We found the structure of typical
singularities appearing in the plasma particle density during the wave
breaking. 
The singularity in the electron density, moving along with the wake wave,
acts as a flying relativistic mirror for the counter-propagating
electromagnetic radiation, reflecting the light and up-shifting its
frequency. Such the reflection results in generation of an ultra-intense
electromagnetic radiation. We found the coefficients of the reflection of an
electromagnetic wave at the singularities (crests) of the electron density
piled up in the most typical regimes of strongly nonlinear wave breaking in
collisionless plasmas. Although the reflection coefficients are small, they
are not exponentially small, as it is usual in the geometric-optic
approximation of the over-barrier reflection from the electron density
modulations, applicable for the wake wave far below the breaking threshold.
The efficiency of the photon reflection can be substatially increased by
using the regimes providing high order sigularity formation (see Eq. (\ref%
{Ralpha}) or by realizing the laser pulse reflection from several subsequent
density singularities, which can be naturally produced in the long enough
wake wave. We show that in the multiple-mirror configuration, the energy of
the counter-propagating laser pulse reflected back and pumped by the wake
wave can be increased by the factor of $4\gamma _{M}^{2}$. 
This paves a way towards producing high efficiency compact sources of hard
electromagnetic radiation.

\section*{Asknowledgments}

This work was supported in part by the Japanese Ministry of Education,
Science, Sports and Culture, Grant-in-Aid for Scientific Research (A),
20244065, 2008, and by the Special Coordination Fund (SCF) for Promoting
Science and Technology commissioned by the Ministry of Education, Culture,
Sports, Science and Technology (MEXT) of Japan. One of the authors (V. A.
P.) gratefully acknowledges the hospitality of Kansai Photon Science
Institute, JAEA.

\section*{Appendix A}

\section*{Plasma flow breaking in the ponderomotive ion acceleration}

Here we consider another important example when one can neglect the space
charge effects. The quasineutral plasma dynamics is provided by the
electromagnetic wave interaction with the electron-ion plasmas in the limit
of extremely high light intensity in the so-called "radiation pressure
dominated regime" (for details see Refs. \cite{RadF,TZE,FB}). In the
quasineutal plasma motion, the longitudinal components of the electron and
ion velocities are equal to each other, i.~e. the longitudinal component of
the electron momentum is much less than the ion momentum. Due to
conservation of the generalized momentum, the transverse components of the
electron and ion momenta are determined by the interaction with the
electromagnetic wave and are equal to $m_{e}ca({\mathrm{X}})$ and $m_{i}c\mu
a({\mathrm{X}})$, respectively, where $\mu =m_{e}/m_{i}$ is the
electron-to-ion mass ratio. It is easy to show that in this model instead of
the integral (\ref{el-int}) we have the following conservation law 
\begin{equation}
\gamma_i +\mu \gamma_e -\left( \mu \frac{p_e}{m_{e}c}+\frac{p_i}{m_{i}c}%
\right) \mathrm{\beta _{ph}}=\mu h+H.  \tag{A.1}  \label{el-ion-int}
\end{equation}%
Here $p_e$, $\gamma_e$ and $p_i$, $\gamma_i =\sqrt{1+\left( \mu a({\mathrm{X}%
})\right) ^{2}+\left(p_i/m_{i}c\right) ^{2}}$ are momentum and the Lorentz
factor for the electron and ion, respectively. In the right hand side part
of Eq. (\ref{el-ion-int}) the constant $H$ is given by $H=\gamma_{i0}-\left(
p_{i0}/m_{i}c\right) \mathrm{\beta _{ph}}$.

Using the quasineutrality condition, $p_i/m_{i}\gamma_i =p_e/m_{e}\gamma_e $%
, and smallness of the parameter $\mu $, we reduce Eq. (\ref{el-ion-int}) to 
\begin{equation}
\sqrt{1+\left( \frac{p_i}{m_{i}c}\right) ^{2}}-\frac{p_i}{m_{i}c}\mathrm{%
\beta _{ph}}=1+\mu \left( 1-\sqrt{1+a^{2}({\mathrm{X}})}\right) ,  \tag{A.2}
\end{equation}%
and find the ion momentum 
\begin{equation}
p_i=m_{i}c\frac{\tilde{H}({\mathrm{X}})\mathrm{\beta _{ph}}-\sqrt{\tilde{H}%
^{2}({\mathrm{X}})-1}}{\sqrt{1-\mathrm{\beta _{ph}^{2}}}}.  \tag{A.3}
\end{equation}%
Here $\tilde{H}({\mathrm{X}})$ is a function given by the expression 
\begin{equation}
\tilde{H}({\mathrm{X}})=\frac{1+\mu \left( 1-\sqrt{1+a^{2}({\mathrm{X}})}%
\right) }{\sqrt{1-\mathrm{\beta _{ph}^{2}}}}.  \tag{A.4}
\end{equation}%
If the laser pulse amplitude is so large that $a>\mu ^{-1}\sqrt{(1+\mu
)^{2}-\mu ^{2}}$, the function $\tilde{H}({\mathrm{X}})$ vanishes at ${%
\mathrm{X=X}}_{br}$ and the ion momentum and energy become equal to $m_{i}c%
\mathrm{\beta _{ph}}/\sqrt{1-\mathrm{\beta _{ph}^{2}}}$ and $m_{i}c^{2}/%
\sqrt{1-\mathrm{\beta _{ph}^{2}}}$. The plasma layer bounces at this point
and then moves in the same direction as the electromagnetic pulse. Its
momentum increases from $m_{i}c\mathrm{\beta _{ph}}/\sqrt{1-\mathrm{\beta
_{ph}^{2}}}$ at ${\mathrm{X=X}}_{br}$ to $m_{i}c\sqrt{(1+\mathrm{\beta _{ph})%
}/(1-\mathrm{\beta _{ph}})}$ at ${\mathrm{X}}{\mathrm{\rightarrow \infty }}$%
. More detailed analysis of this regime of the ion acceleration will be
presented elsewhere. 

\section*{Appendix B}

\section*{Geometric properties of caustics in plasma flows}

We found above three kinds of the singularity formed due to the nonlinear
wake wave breaking: the particle density depends on the coordinate as $n(%
\mathrm{X})\propto {{\mathrm{X}}}^{-\alpha }$, where the exponent $\alpha $
is equal to $1/2$, $2/3$, and $1$. This exponent represents the measure of
the singularity strength. Here we present a generic description of nonlinear
wave breaking, based on the geometric properties of caustics in the plasma
flow. It is convenient to use the Lagrange variables, $(x_{0},t)$ , related
to the Euler coordinates $(x,t)$ as 
\begin{equation}
x=x_{0}+\xi (x_{0},t)  \tag{B.1}
\end{equation}%
with $\xi (x_{0},t)$ being the electron displacement and $v=\partial _{t}\xi 
$ . As known, the solution to the continuity equation in the Lagrange
coordinates is given by the expression%
\begin{equation}
n(x_{0},t)=\frac{{n_{0}}}{{J(x_{0},t)}}  \tag{B.2}
\end{equation}%
for the particle density, where%
\begin{equation}
J(x_{0},t)=\left \vert \frac{{\partial (x,t)}}{{\partial (x_{0},t)}}\right
\vert =\left \vert 1+\partial _{x_{0}}\xi \right \vert  \tag{B.3}
\end{equation}%
is the Jacobian of transformation from the Lagrange to the Euler
coordinates. The transformation has a singularity at the point where the
Jacobian vanishes, $J(x_{0},t)\rightarrow 0$ , which is equivalent to the
condition $\partial _{x_{0}}\xi \rightarrow -1$ . This singularity
corresponds to the wave breaking, when the electron density and the velocity
(momentum) gradient become infinite.

In the vicinity of the singularity, where the Lagrange coordinate is equal
to $x_{0}=x_{br}$ , i.e for $x_{0}=x_{br}+\delta x_{0}$ with $\delta
x_{0}\ll 1$ , we can expand the dependence of the Euler coordinate on $%
x_{br} $ and $\delta x_{0}$ over $\delta x_{0}$ 
\begin{equation}
x=x_{br}+\delta x_{0}+\xi (x_{br}+\delta x_{0})=
\nonumber
\vspace*{-3ex}
\end{equation}%
\begin{equation}
=x_{br}+\left( \frac{\xi
_{br}^{\prime \prime }}{2}\right) \delta x_{0}^{2}+\ldots \left( \frac{\xi
^{(j)}}{j!}\right) \delta x_{0}^{j}+\ldots  \tag{B.4}  \label{xxbr}
\end{equation}
Here the subscript 'br' means that the value of a function is taken at $%
x_{0}=x_{br}$ , the prime denotes differentiation with respect to $x_{0}$ ,
and we take into account that $1+\xi _{br}^{\prime }=0$ . Expanding the
Jacobian $J(x_{0},t)$ in series of $\delta x_{0}$ we obtain%
\begin{equation}
J=|\xi _{br}^{\prime \prime }|\delta x_{0}+\ldots +\left \vert \frac{\xi
_{br}^{(j)}}{(j-1)!}\right \vert \delta x_{0}^{j-1}+\ldots  \tag{B.5}
\end{equation}

If the second derivative $\xi _{br}^{\prime \prime }$ does not vanish, we
obtain the following relationship between the Lagrange and Euler
coordinates: $\delta x_{0}\approx ({2}(x-x_{br})/{\xi _{br}^{\prime \prime }}%
)^{1/2}$. For the electron density we have 
\begin{equation}
n\approx \frac{n_{0}}{\left( 2\xi _{br}^{\prime \prime }\right) ^{1/2}}%
(x-x_{br})^{-1/2}.  \tag{B.6}
\end{equation}%
This case corresponds to the regime described by Eq. (\ref{nx1/2}).

In the case when the second derivative of the displacement vanishes at the
singularity, $\xi _{br}^{\prime \prime }=0$, and the third derivative is
nonzero, $\xi _{br}^{\prime \prime \prime }\neq 0$, we find $\delta
x_{0}\approx ({6}(x-x_{br})/{\xi _{br}^{\prime \prime \prime }})^{1/3} $ and 
\begin{equation}
n\approx n_{0}\left( \frac{2}{9\xi _{br}^{\prime \prime \prime }}\right)
^{1/3}(x-x_{br})^{-2/3}.  \tag{B.7}
\end{equation}%
In this case we recover the result obtained above: the singularity $n\propto 
{\Delta \mathrm{X}}^{-2/3}$ in the wake wave breaking described by Eqs. (\ref%
{v2/3},\ref{n2/3}).

If all the derivatives of the displacement $\xi $ with respect to the
variable $x_{0}$ of the order less than $m$ vanish, we obtain from Eq. (\ref%
{xxbr}) the relationship between $x$ and $\delta x_{0}$: 
\begin{equation}
x=x_{br}+\frac{{\xi _{br}^{(m)}}}{{m!}}\delta x_{0}^{m}+\ldots .  \tag{B.8}
\end{equation}%
It yields for the Jacobian 
\begin{equation}
J=m\left( \frac{\xi _{br}^{(m)}}{m!}\right) ^{1/m}(x-x_{br})^{\alpha _{m}} 
\tag{B.9}
\end{equation}%
and for the electron density%
\begin{equation}
n=n_{0}\frac{1}{m}\left( \frac{m!}{\xi _{br}^{(m)}}\right)
^{1/m}(x-x_{br})^{-\alpha _{m}},  \tag{B.10}  \label{n1/m}
\end{equation}%
where the exponent $\alpha _{m}$ is equal to 
\begin{equation}
\alpha _{m}=\frac{m-1}{m}  \tag{B.11}
\end{equation}%
with $m\geq 2$.

For $m\gg 1$ we use the Stirling formula for the factorial $m!=\sqrt{2\pi }%
m^{m+1/2}\exp (-m+r_{m})$ with $1/(12m+1)<r_{m}<1/12m$. Representing the
derivative $\xi _{br}^{(m)}$ as $\xi _{br}^{(m)}= \mathrm{const}%
\lambda_{wf}^{-m}$, where $\lambda_{wf}$ is the wake field wavelength, we
find from Eq. (\ref{n1/m}) that in the limit $m\rightarrow \infty $ the
density distribution is described by 
\begin{equation}
n={\boldsymbol{e}^{-1}} \frac{n_{0}\lambda _{wf}}{x-x_{br}}.  \tag{B.12}
\end{equation}%
Here $\boldsymbol{e}=2.718...$ is base of the natural logarithm (the Napier
constant). This dependence of the electron density on the coordinate $%
(x-x_{br})$ corresponds to Eq. (\ref{n(1/x)}).


\begin{thebibliography}{99}
\bibitem{rel-beam}
K. Landecker, Phys. Rev. \textbf{86}, 852 (1952);
L. A. Ostrovskii, Sov. Phys. Usp. \textbf{18}, 452 (1976);
V. L. Granatstein, et al. 
Phys. Rev. A \textbf{14}, 1194.(1967);
J. A. Pasour, V. L. Granatstein, and R. K. Parker, Phys. Rev. A \textbf{16}, 2441 (1977).

\bibitem{Photon acceleration}
C. Wilks, et al. 
Phys. Rev. Lett. \textbf{62}, 2600 (1989); S.
V. Bulanov and A. S. Sakharov, JETP Lett. \textbf{54}, 203(1991);
C. W. Siders, et al. 
Phys. Rev. Lett. \textbf{76}, 3570 (1996);
J. M. Dias, et al. 
Phys. Rev. Lett. \textbf{78}, 4773 (1997);
J. T. Mendonca, Photon Acceleration in Plasmas (Inst. Phys. Publ., Bristol, 2001);
C. D. Murphy, et al. 
Phys. Plasmas \textbf{13}, 033108 (2006).

\bibitem{KrSoob}
S. V. Bulanov, et al., Kratk. Soobshch. Fiz. ANSSSR \textbf{6}, 9 (1991), in Russian;
S. V. Bulanov, et al. 
in: Reviews of Plasma Physics. Volume: 22, edited by V. D. Shafranov, (Kluwer Academic / Plenum Publishers, New York, 2001), p. 227.

\bibitem{Light intensification}
S. V. Bulanov, T. Zh. Esirkepov and T. Tajima, Phys. Rev. Lett. \textbf{91}, 085001 (2003).

\bibitem{Kando-2007}
M. Kando, et al. 
Phys. Rev. Lett. \textbf{99}, 135001 (2007).

\bibitem{Pirozhkov-2007}
A. S. Pirozhkov, et al. 
Plasma Phys. \textbf{14}, 123106 (2007).

\bibitem{Ioniz}
V. I. Semenova, Sov. Radiophys. Quantum Electron. \textbf{10}, 599 (1967);
M. Lampe, E. Ott and J.H. Walker. Phys. Fluids \textbf{21}, 42(1978);
W. B. Mori, Phys. Rev. A \textbf{44}, 5118 (1991);
R. L. Savage, Jr. et al., Phys. Rev. Lett. \textbf{68}, 946 (1992);
W. Yu, et al. 
J. Phys. D \textbf{26}, 2093 (1993);
X. Feng and S. Lee, Optics Communications \textbf{136}, 385 (1997).

\bibitem{RelSol}
S. S. Bulanov, et al. 
Phys. Rev. E \textbf{73}, 036408 (2006).

\bibitem{OscMir}
S. V. Bulanov, N. M. Naumova and F. Pegoraro, Phys. Plasmas \textbf{1}, 745 (1994);
R. Lichters, J. Meyer-ter-Vehn and A. M. Pukhov, Phys. Plasmas \textbf{3}, 3425 (1996);
V. A. Vshivkov, et al. 
Phys. Plasmas \textbf{5}, 2727 (1998);
A. S. Pirozhkov, et al. 
Phys. Plasmas \textbf{13}, 013107 (2006).

\bibitem{Cher}
V. V. Kulagin, et al. 
Phys. Plasmas \textbf{14}, 113101 (2007).

\bibitem{Naumova}
N. M. Naumova, et al. 
Phys. Rev. Lett. \textbf{92}, 063902 (2004);
N. M. Naumova, J. Nees, and G. Mourou, Phys. Plasmas \textbf{12}, 056707 (2005);
N. M. Naumova, et al. 
New J. Phys. \textbf{10,} 025022 (2008).

\bibitem{1022}
V. Yanovsky, et al. 
Optics Express \textbf{16}, 2109 (2008).

\bibitem{RadF}
Ya. B. Zel'dovich and A. F. Illarionov, Sov. Phys. JETP 34, 467 (1971);
A. D. Steiger and C. H. Woods, Phys. Rev. E 5, 1467 (1972);
Ya. B. Zel'dovich, Sov. Phys. Usp. 18, 79 (1975);
A. G. Zhidkov, et al. 
Phys. Rev. Lett. \textbf{88}, 185002 (2002);
J. Koga, T. Zh. Esirkepov and S. V. Bulanov, Phys. Plasmas \textbf{12}, 093106 (2005).

\bibitem{BEKT}
S. V. Bulanov, et al. 
Plasma Phys. Rep. \textbf{30}, 221 (2004).

\bibitem{H-E}
W. Heisenberg and H. Z. Euler, Z. Phys. \textbf{98}, 714 (1936);
J. Schwinger, Phys. Rev.\textbf{\ 82}, 664 (1951);
E. Brezin and C. Itzykson, Phys. Rev. D \textbf{2}, 1191 (1970);
N. B. Narozhny and A. I. Nikishov, Sov. Phys. JETP \textbf{38}, 427 (1973);
V. S. Popov, JETP \textbf{94}, 1057 (2002);
N. B. Narozhny, et al. 
Phys. Lett. A \textbf{330}, 1 (2004).

\bibitem{MTB}
A. Mourou, T. Tajima and S. V. Bulanov, Rev. Mod. Phys. \textbf{78}, 309 (2006);
M. Marklund and P. K. Shukla, Rev. Mod. Phys. \textbf{78}, 591 (2006);
Y. I. Salamin, et al. 
Phys. Rep. \textbf{427}, 41 (2006).

\bibitem{RCNP}
S. V. Bulanov, et al. 
in: "Proceedings of the International Workshop on Quark Nuclear
Physics", Eds.: J. K. Ahn, M. Fujiwara, T. Hayakawa, et al. 
Pusan National Univesity Press, PUSAN, 2006, p.179.

\bibitem{D-shape}
S. V. Bulanov and A. S. Sakharov, JETP Lett., \textbf{54}, 203 (1991);
S. V. Bulanov, F. Pegoraro and A. M. Pukhov, Phys. Rev. Lett.  \textbf{74}, 710 (1995);
Z.-M. Sheng, et al. 
Phys. Rev. E \textbf{62}, 7258 (2000);
N. H. Matlis, et al. 
Nature Phys. \textbf{2}, 749 (2006);
A. Maksimchuk et al. 
Phys. Plasmas \textbf{15}, 056703 (2008).

\bibitem{Akhiezer-Polovin}
A. I. Akhiezer and R. V. Polovin, Sov. Phys. JETP \textbf{3}, 696 (1956);
O. B. Shiryaev, Phys. Plasmas \textbf{15}, 012308 (2008).

\bibitem{Einstein}
A. Einstein, Ann. Phys. (Leipzig) \textbf{17}, 891 (1905).

\bibitem{Z-R}
Ya. B. Zel'dovich and Yu. P. Raizer, Physics of Shock Waves and High-Temperature Hydrodynamic Phenomena (Academic, New York, 1967).

\bibitem{BBK}
B. B. Kadomtsev, in: Reviews of Plasma Physics. Volume: 22,
Ed.: V. D. Shafranov, (Kluwer Academic / Plenum Publishers, New York, 2001), p. 1.

\bibitem{PhMix}
S. V. Bulanov, L. M. Kovrizhnykh and A. S. Sakharov, Physics Reports \textbf{186}, 1 (1990);
S. V. Bulanov, et al. 
Phys. Rev. E \textbf{58}, R5257 (1998);
G. Lehmann, E. W. Laedke and K. H. Spatschek, Phys. Plasmas\textbf{\ 14}, 103109 (2007);
C. G. R. Geddes, et al. 
Phys. Rev. Lett. \textbf{100}, 215004 (2008);
A. V. Brantov, et al. 
Phys. Plasmas, in press (2008).

\bibitem{TZE}
T. Esirkepov, et al. 
Phys. Rev. Lett. \textbf{92}, 175003 (2004);
F. Pegoraro and S. V. Bulanov, Phys. Rev. Lett. 99, 065002 (2007);
O. Klimo, et al. 
Phys. Rev. STAB \textbf{11}, 031301 (2008);
X. Q. Yan, et al. 
Phys. Rev. Lett. \textbf{100}, 135003 (2008).

\bibitem{FB}
D. Farina and S. V. Bulanov, Plasma Phys. Control. Fusion  \textbf{47}, A73 (2005).

\bibitem{Liu-Ros}
M. N. Rosenbluth and C. S. Liu, Phys. Rev. Lett. \textbf{29}, 701 (1972).

\bibitem{T-D}
T. Tajima and J. Dawson, Phys. Rev. Lett. \textbf{43}, 267 (1979).

\bibitem{GK}
L. M. Gorbunov and V. I. Kirsanov, Sov. Phys. JETP \textbf{66}, 290 (1987).

\bibitem{Esarey-IEEE}
E. Esarey,  et al. 
IEEE Trans. Plasma Sci. \textbf{24}, 252 (1996).

\bibitem{Esarey}
P. Sprangle,  et al. 
Optics Communications \textbf{124}, 69 (1996)

\bibitem{Chen}
P. Chen, et al. 
Phys. Rev. Lett. \textbf{54}, 693 (1985).

\bibitem{EBYT}
T. Esirkepov, et al. 
Phys. Rev. Lett. \textbf{96}, 014803 (2006).

\bibitem{A-S}
M. Abramowitz and I. A. Stegun,
Handbook of Mathematical Functions with Formulas, Graphs, and Mathematical Tables
(Dover, New York, 1964).

\bibitem{Budden}
K. G. Budden, Propagation of Radio Waves (Cambridge University Press, Cambridge, 1988).

\bibitem{Stix}
T. H. Stix, Waves in Plasmas (American Institute of Physics, New York, 1992).

\end{thebibliography}
\end{document}